\documentclass[12pt]{article}
\usepackage[utf8]{inputenc}
\usepackage[margin=1in]{geometry}
\usepackage{amsmath}
\usepackage{amssymb}
\usepackage{graphicx}
\usepackage{natbib}
\usepackage{hyperref}
\usepackage{booktabs}
\usepackage{array}
\usepackage{caption}

\title{Tournament-Based Performance Evaluation and Systematic Misallocation: \\
Why Forced Ranking Systems Produce Random Outcomes}

\author{Jeremy McEntire}
\date{December 6, 2025}

\begin{document}

\maketitle

\begin{abstract}
Tournament-based compensation schemes with forced distributions represent a widely adopted class of relative performance evaluation mechanisms in technology and corporate environments. These systems mandate within-team ranking and fixed distributional requirements (e.g., bottom 15\% terminated, top 15\% promoted), ostensibly to resolve principal-agent problems through mandatory differentiation. We demonstrate through agent-based simulation that this mechanism produces systematic classification errors independent of implementation quality. With 994 engineers across 142 teams of 7, random team assignment yields 32\% error in termination and promotion decisions, misclassifying employees purely through composition variance. Under realistic conditions reflecting differential managerial capability, error rates reach 53\%, with false positives and false negatives each exceeding correct classifications. Cross-team calibration (often proposed as remedy) transforms evaluation into influence contests where persuasive managers secure promotions independent of merit. Multi-period dynamics produce adverse selection as employees observe random outcomes, driving risk-averse behavior and high-performer exit. The efficient solution (delegating judgment to managers with hierarchical accountability) cannot be formalized within the legal and coordination constraints that necessitated forced ranking. We conclude that this evaluation mechanism persists not through incentive alignment but through satisfying demands for demonstrable process despite producing outcomes indistinguishable from random allocation. This demonstrates how formalization intended to reduce agency costs structurally increases allocation error.
\end{abstract}


\section{Introduction}

\subsection{The Promise of Forced Ranking}

Forced ranking (also known as stack ranking, vitality curves, or rank-and-yank) represents a tournament-based compensation scheme where managers evaluate employees through relative performance assessment within teams, assigning them to predetermined categories following a fixed distribution (e.g., top 15\%, middle 70\%, bottom 15\%). This mechanism was popularized by Jack Welch at General Electric in the 1980s and subsequently adopted across technology companies, consulting firms, and financial services as a solution to principal-agent problems in performance evaluation \citep{scullen2005forced, stewart2010forced}.

Tournament theory provides the intellectual foundation for such systems. \citet{lazear1981rank} demonstrated that rank-order tournaments can function as optimal labor contracts when individual output is difficult to measure but relative performance is observable. The tournament mechanism creates incentives through prize spreads between ranks rather than through piece-rate compensation tied to absolute output. By evaluating employees relative to peers rather than against uncertain absolute standards, organizations theoretically reduce measurement costs while maintaining incentive intensity.

The theoretical appeal rests on several mechanisms. By requiring differentiation within teams and imposing distributional constraints, forced ranking purports to:

\begin{itemize}
\item Eliminate ratchet effects and grade inflation by preventing convergence to uniform high ratings
\item Reduce information asymmetries between principals (boards, shareholders) and agents (managers conducting evaluations)
\item Create accountability through quantified, comparable metrics that traverse organizational hierarchies
\item Identify low-productivity workers systematically rather than allowing managers to avoid costly dismissal decisions
\item Ensure tournament prizes reach top performers in every unit, maintaining incentive intensity across the organization
\item Provide legal defensibility through documented processes satisfying fiduciary requirements for demonstrable soundness
\end{itemize}

The mechanism appears to resolve fundamental information problems in employment relationships. It transforms distributed, contextual judgment into standardized rankings, applies consistent criteria across organizational units, and produces documented decisions defensible to boards, regulators, and in litigation. This formalization ostensibly reduces agency costs by constraining managerial discretion while maintaining performance incentives.

\subsection{The Practice and Its Critics}

Despite this theoretical appeal, forced ranking has generated sustained criticism from practitioners, organizational scholars, and former executives at companies that implemented it. Recent systematic reviews following PRISMA procedures have analyzed the accumulated evidence, finding that while differentiation within forced distribution systems may initially elicit positive performance reactions, serious injustice perceptions result in counterproductive behaviors over time, particularly under high task interdependence \citep{wijayanti2023systematic}.

Microsoft abandoned the system in November 2013 after internal assessment concluded it fostered toxic competition, discouraged collaboration, and contributed to strategic missteps \citep{eichenwald2012microsoft}. General Electric quietly moved away from it beginning around 2005, with a complete overhaul in 2015 \citep{microsoft2013}. Yahoo's implementation under Marissa Mayer drew immediate backlash. Academic research has documented negative effects on teamwork \citep{pfeffer2001fighting, loberg2021forced}, gaming behavior \citep{scullen2005forced}, increased employee stress \citep{cardinaels2021forced}, and challenges to organizational justice \citep{moon2016precarious}.

Contemporary performance management research has identified that current systems fail because they focus too narrowly on individual outcomes rather than taking a systems perspective, particularly neglecting how elements interact across organizational levels \citep{schleicher2018putting}. A 30-year integrative review found that research has overemphasized reactions and ratings while neglecting learning outcomes, managerial behaviors, and unit-level impacts \citep{schleicher2019evaluating}.

Yet the practice persists, particularly in technology companies and professional services firms. Why? The conventional explanation focuses on managerial incompetence or cultural problems: CEOs who worship metrics, HR departments that prioritize process over outcomes, managers who lack courage to make difficult decisions. These critiques suggest forced ranking fails through poor implementation rather than structural deficiency.

Yet even as high-profile abandonments mount, 2025 has witnessed a concerning resurgence. Recent industry analyses document that some technology firms, under economic pressure and leadership transitions, are tentatively reviving forced ranking under rebranded terminologies, framed as ``rigor'' or ``performance calibration'' rather than ``rank-and-yank'' \citep{kornferry2025revival}. These revivals reproduce familiar pathologies: increased voluntary turnover among high performers, exacerbated team dysfunction, and renewed complaints about arbitrary evaluation. The pattern suggests that forced ranking's appeal to demonstrable soundness remains potent despite accumulated evidence of its destructive effects.

\subsection{This Paper's Contribution}

We demonstrate that forced ranking constitutes a structurally deficient incentive mechanism that produces systematic misallocation independent of implementation quality, managerial capability, or organizational culture. Using agent-based simulation, we show that tournament-based evaluation with forced distributions generates classification errors inherent to the mechanism's information structure. Even under idealized conditions (random team assignment eliminating selection bias, perfect managerial observation within teams, no measurement error in assessing talent), the mechanism misclassifies approximately one-third of employees. Under realistic conditions where team quality varies (reflecting differential managerial capability in attracting, developing, and retaining talent), error rates exceed 50\%, meaning incorrect allocative decisions outnumber correct ones.

These errors constitute not implementation failures but mathematical necessities emerging from a fundamental information problem: evaluating a global population through local frames. Forced ranking rests on an invalid statistical assumption that small teams represent unbiased samples of the organization's global talent distribution. This exemplifies the \textbf{small sample fallacy}, where small groups are erroneously expected to mirror population properties \citep{tversky1971law}. Team size creates high composition variance that generates classification errors no amount of procedural refinement can eliminate. Counter-intuitively, this allocation failure \textit{worsens} with better management. As capable managers attract and develop talent, they create high-performing teams; this non-random sorting amplifies statistical error and ensures the mechanism punishes effective management most severely. Real-world selection pressures compound this structural deficiency.

We further demonstrate that commonly proposed remedial mechanisms fail or create second-order distortions:

\begin{itemize}
\item Cross-team calibration transforms evaluation from information aggregation into influence contests, where persuasive managers secure allocations independent of subordinate productivity
\item Absolute performance standards require global information that no organizational actor possesses, recreating the information problem formalization was meant to solve
\item Multi-period dynamics produce adverse selection as high-productivity workers observe random allocation and exit, while risk-averse workers accumulate
\end{itemize}

The paper proceeds as follows. Section 2 develops the theoretical framework connecting tournament-based evaluation to information structure problems in organizational economics. Section 3 describes simulation methodology. Section 4 presents results under random and non-random team assignment. Section 5 analyzes why proposed remedial mechanisms fail. Section 6 examines multi-period dynamics and adverse selection. Section 7 discusses why the efficient solution---delegating judgment to managers with hierarchical accountability---cannot be formalized within existing institutional constraints. Section 8 concludes with implications for incentive mechanism design and personnel economics.

\section{Theoretical Framework: Information Structure and Allocation Mechanisms}

\subsection{The Information Problem in Performance Evaluation}

Performance evaluation in organizations confronts a fundamental information asymmetry. Accurate assessment requires detailed knowledge of individual contributions, project difficulty, team dynamics, and comparative productivity. This information is distributed: direct managers possess granular visibility into their teams through repeated observation, but no single organizational actor commands firm-wide perspective enabling global comparisons. This creates the classic principal-agent problem where those with evaluation authority (managers) possess superior information compared to those requiring aggregated assessments (executives, boards, shareholders).

Organizations operating at scale simultaneously face legal, coordination, and accountability pressures demanding standardization. The business judgment rule protects directors who demonstrate informed, documented decision processes \citep{smith1985, badawi2023business, sharfman2017importance}. Human resources functions must defend compensation and termination decisions to regulators, auditors, and potential litigants. Boards require evidence that talent allocation mechanisms operate systematically rather than arbitrarily.

Tournament-based forced ranking resolves this tension through formalization: transforming distributed, contextual judgment into standardized, quantified rankings. This mechanism reflects what organizational scholars identify as ``coercive'' formalization functioning as managerial control, distinct from ``enabling'' formalization that supports task performance \citep{adler1996two}. By imposing distributional constraints (e.g., every team must designate bottom 15\% for termination), forced ranking creates:

\begin{itemize}
\item Comparability across organizational units enabling aggregated assessment
\item Documentation of systematic process satisfying legal and fiduciary requirements
\item Defensibility against litigation claims of bias or arbitrariness
\item Accountability through mandatory differentiation constraining managerial discretion
\end{itemize}

This formalization ostensibly eliminates discretionary judgment that creates legal exposure. Rather than managers claiming ``no one on my team merits termination,'' forced ranking mandates binary classification: someone receives bottom ranking regardless of absolute performance level. The mechanism appears to reduce both agency costs and litigation risk through procedural constraint.

\subsection{The Allocation Problem: Local Information, Global Decisions}

Formalization creates what we term an \emph{allocation mechanism with mismatched information structure}. Tournament-based forced ranking operates within team boundaries: each manager ranks subordinates relative to teammates and applies distributional requirements locally. However, the allocative decisions these local rankings drive---termination, promotion, compensation---have firm-wide implications determining who remains employed and who advances.

The local information structure can validate only within-team comparisons. A manager can accurately determine that Employee A exhibits higher productivity than Employee B on their team. Forced ranking, however, requires inferring from this local observation a global claim: that the bottom-ranked employee on Team 1 demonstrates lower productivity than the bottom-ranked employee on Team 2. This inference holds only if teams constitute representative, unbiased samples of the firm's global productivity distribution.

They do not. Team composition varies through multiple selection mechanisms:

\begin{itemize}
\item \textbf{Hiring:} Capable managers attract higher-quality candidates; less effective managers face adverse selection in recruitment
\item \textbf{Development:} Differential managerial capability produces heterogeneous within-team productivity growth
\item \textbf{Retention:} High-productivity workers exit teams with poor management; lower-productivity workers demonstrate less mobility
\item \textbf{Functional assignment:} Workers with specialized human capital cluster in particular organizational units
\item \textbf{Growth dynamics:} Successful teams expand and recruit; struggling teams contract through attrition
\end{itemize}

Research on team composition demonstrates how member characteristics aggregate to influence collective outcomes \citep{stewart2006meta, bell2007deep}. Meta-analytic evidence shows that aggregated individual ability (especially general mental ability) and personality dispositions significantly relate to team performance, with team minimum and maximum ability particularly influential \citep{bell2007deep}. This creates fundamental challenges for evaluation mechanisms that ignore team quality heterogeneity.

Recent empirical work reinforces these concerns. Field studies find that restricting top performance ratings---a common implementation of forced distributions---drives voluntary turnover among precisely the high-achieving employees organizations seek to retain \citep{cornell2025rankings}. Systems-theory analyses of forced distribution in practice reveal that over 50\% of rating variance stems from rater biases rather than actual performance differences, with these biases systematically disadvantaging minorities and employees in small or dynamic teams \citep{hrdm2025systems}. Bell curves, rather than resolving evaluation challenges, entrench inequities while creating procedural facades that obscure rather than illuminate actual capability differences.

Even random assignment (our idealized baseline) produces variance. With 994 engineers drawn from a standard normal distribution assigned to 142 teams of 7, team means vary by construction. Pure chance creates teams where the worst performer is in the global top quartile and teams where the best performer is in the global bottom quartile.

\subsection{Mechanism Design Under Institutional Constraints}

This allocation problem is not unique to forced ranking. It exemplifies a general tension in organizational governance: formalization creates demonstrable process soundness while destroying information aggregation capacity. Organizations under fiduciary duty require documented, defensible procedures (what corporate governance scholars term procedural rationality). However, demonstrable procedural soundness mandates evaluation within formal constraints: metrics, standardized processes, quantified criteria. Information existing outside these formal constraints cannot be validated through mechanisms designed to operate within them.

Tournament-based forced ranking represents formalized evaluation in its purest form:

\begin{itemize}
\item It eliminates variance (all managers apply identical distributional requirements)
\item It creates defensibility (documented rankings, standardized process)
\item It operates within defined boundaries (team-level information structure)
\item It systematically excludes external information (global productivity distribution)
\end{itemize}

The result is a mechanism that appears procedurally rigorous yet produces allocative outcomes structurally inferior to discretionary judgment-based evaluation. Like other organizational formalizations, tournament-based forced ranking fails not through poor implementation but because its information structure cannot access the data necessary for efficient allocation decisions.

\subsection{Why Efficient Mechanisms Cannot Be Implemented}

The efficient solution (evaluate employees against global productivity standards using firm-wide information) confronts insurmountable implementation barriers rooted in information costs, coordination costs, and institutional constraints:

\textbf{Information constraint:} No single actor possesses sufficient information to rank 994 engineers. Evaluation must be distributed to managers possessing localized knowledge through repeated observation. Distributed evaluation, however, reintroduces the allocation problem: each manager operates within their local information structure, and aggregating local assessments cannot recover the global productivity distribution they individually lack access to.

\textbf{Coordination constraint:} Cross-team calibration requires managers to share information, negotiate relative rankings, and reach consensus. This creates second-order mechanism design failures: evaluation transforms into influence contests, persuasive managers secure disproportionate allocations, empire-building becomes individually rational. Allocative error does not decrease; it becomes more difficult to detect and correct.

\textbf{Institutional constraint:} Discretionary judgment-based evaluation (managers determining termination based on contextual assessment) cannot be documented satisfying business judgment rule requirements \citep{smith2015modern}. Managers cannot credibly demonstrate to legal, human resources, or boards that their team genuinely exhibits high productivity warranting zero terminations. The absence of quantifiable procedural constraints creates litigation exposure and fiduciary risk.

The mechanism design trap closes: tournament-based forced ranking exhibits structural allocation failures, yet the alternative (discretionary managerial judgment) creates institutional risk and imposes prohibitive coordination costs. Organizations adopt forced ranking not through managerial incompetence but because it satisfies institutional requirements for demonstrable process despite producing demonstrably inefficient allocative outcomes.

The simulation in Section 3 quantifies this mechanism's allocative error magnitude.

\section{Methodology: Simulation Design}

\subsection{Agent-Based Simulation Structure}

We construct an agent-based simulation modeling a technology company with 994 software engineers distributed across 142 teams of 7 engineers each. Agent-based simulation has become an established methodology in organizational research for examining complex organizational processes and behaviors \citep{fioretti2013agent, harrison2007simulation, gomez2017agent}. The simulation proceeds as follows:

\textbf{Step 1: Talent Generation}

994 individual talents drawn from standard normal distribution: 
\begin{equation}
\text{Talent}_i \sim N(0, 1)
\end{equation}

This creates a bell curve where approximately 68\% of employees are within one standard deviation of the mean, with tails representing exceptional and poor performers.

\textbf{Step 2: Team Assignment}

We implement two assignment variants:

\emph{Random Assignment (Baseline):} Engineers randomly assigned to teams with no constraints. This represents the best-case scenario for forced ranking---no hiring bias, no managerial quality differences, no favoritism. Team composition variance exists purely due to sampling variation.

\emph{Biased Assignment (Realistic):} Team quality varies to simulate differential managerial capability. Implementation:
\begin{itemize}
\item Draw 142 team means from $N(0, 0.7)$
\item For each team with mean $\mu_{\text{team}}$, draw 7 members from $N(\mu_{\text{team}}, 0.714)$
\item Overall distribution maintains $N(0, 1)$ by construction ($0.7^2 + 0.714^2 \approx 1$)
\end{itemize}

This clusters high performers in ``strong'' teams (good managers who attract/develop talent) and low performers in ``weak'' teams (poor managers).

\textbf{Step 3: Ground Truth Identification}
\begin{itemize}
\item Rank all 994 engineers by true talent globally
\item Identify true bottom 15\% ($\approx$149 engineers) who should be terminated
\item Identify true top 15\% ($\approx$149 engineers) who should be promoted
\end{itemize}

\textbf{Step 4: Forced Ranking Application}

Within each team:
\begin{itemize}
\item Rank 7 members by talent
\item Label bottom $\approx$15\% (typically 1 per team) for termination: $\lfloor 142 \times 0.15 \rfloor = 21$ terminations per team $\rightarrow$ $\approx$142 total
\item Label top $\approx$15\% (typically 1 per team) for promotion: similarly $\approx$142 promotions
\end{itemize}

\textbf{Step 5: Classification Error Measurement}

Compare forced ranking outcomes to ground truth:

\emph{Terminations:}
\begin{itemize}
\item Correct: Fired AND in true global bottom 15\%
\item False Positive: Fired but NOT in true global bottom 15\%
\item False Negative: In true global bottom 15\% but NOT fired
\end{itemize}

\emph{Promotions:}
\begin{itemize}
\item Correct: Promoted AND in true global top 15\%
\item False Positive: Promoted but NOT in true global top 15\%
\item False Negative: In true global top 15\% but NOT promoted
\end{itemize}

Error Rate: $\frac{\text{False Positives}}{\text{Total Labeled}}$ (proportion of labeled employees incorrectly classified)

\subsection{Implementation Details}

Simulation implemented in Python using NumPy for numerical operations. For each scenario (random, biased), we run 100 independent replications and report mean outcomes with 95\% confidence intervals. Code and data available upon request.

\subsection{Key Assumptions and Limitations}

\textbf{Talent as unidimensional:} We model talent as a single scalar. Real engineers have multidimensional capabilities (coding, design, communication, mentorship), but forced ranking typically reduces these to a single comparative ranking. Our simplification mirrors the practice.

\textbf{Known ground truth:} We assume talent is observable for simulation purposes. In reality, talent is only partially observable, introducing additional measurement error beyond the structural errors we demonstrate \citep{cardinaels2021forced}. Our results thus represent a lower bound on forced ranking's failure rate.

\textbf{Static teams:} Teams do not change composition within a simulation run. Real organizations have turnover, transfers, and growth. We address multi-period dynamics separately in Section 6.

\textbf{No gaming:} Employees and managers do not strategically manipulate rankings. Real forced ranking creates incentives for politics, favoritism, and gaming. Our results exclude these behavioral distortions, again understating real-world error rates.

These simplifying assumptions all favor forced ranking. Real implementations perform worse than our simulations.

\section{Results: Quantifying Classification Errors}

\subsection{Random Assignment: Best-Case Scenario}

Under random team assignment (the most charitable possible conditions), forced ranking produces systematic classification errors.

\begin{table}[h]
\centering
\caption{Random Assignment Results (Mean over 100 simulations)}
\label{tab:random}
\begin{tabular}{lcc}
\toprule
\textbf{Metric} & \textbf{Terminations} & \textbf{Promotions} \\
\midrule
Total Labeled & 142 & 142 \\
Correct Classifications & 97 (68\%) & 96 (68\%) \\
False Positives (Incorrectly Labeled) & 45 (32\%) & 46 (32\%) \\
False Negatives (Missed) & 52 & 53 \\
Error Rate & 32\% & 32\% \\
\bottomrule
\end{tabular}
\end{table}

\textbf{Interpretation:} Of 142 employees forced ranking selects for termination, only 97 (68\%) are actually in the global bottom 15\%. The remaining 45 (32\%) are mid-tier or even strong performers who happened to land on high-performing teams. Conversely, 52 employees who genuinely belong in the bottom 15\% escape termination because they are the ``best of the worst'' on low-performing teams.

Promotions mirror this pattern: 46 employees receive promotions despite not being in the true global top 15\%, while 53 deserving employees are overlooked.

\textbf{Critical observation:} This is not a culture problem, a management problem, or an implementation problem. Under perfect randomization with no bias of any kind, forced ranking produces one-third error rate by construction. The variance in team composition---which cannot be eliminated---creates systematic misclassification.

\textbf{Concrete example:} Consider a team that by chance draws 7 employees all from the 60th--85th percentile (mid-to-strong performers). Forced ranking requires labeling one for termination. That person performs better than 60\% of the company but gets fired because they're surrounded by even stronger peers. Simultaneously, a team drawing from the 15th--40th percentile (weak-to-mid performers) must promote someone. That person underperforms 60\% of the company but gets promoted because they're the ``best'' on a weak team.

This is not a pathology. It is the necessary consequence of applying distributional requirements to non-representative samples.

\subsection{Biased Assignment: Realistic Conditions}

Real organizations do not randomly assign employees to teams. Strong managers attract capable employees; weak managers drive them away. Projects with high visibility and impact draw top talent; legacy maintenance teams do not. Some functions require specialized skills that cluster; others are more general. To model this, we introduce team quality variance.

\begin{table}[h]
\centering
\caption{Biased Assignment Results (Mean over 100 simulations)}
\label{tab:biased}
\begin{tabular}{lcc}
\toprule
\textbf{Metric} & \textbf{Terminations} & \textbf{Promotions} \\
\midrule
Total Labeled & 142 & 142 \\
Correct Classifications & 66 (46\%) & 66 (46\%) \\
False Positives (Incorrectly Labeled) & 76 (53\%) & 76 (53\%) \\
False Negatives (Missed) & 83 & 83 \\
Error Rate & 53\% & 53\% \\
\bottomrule
\end{tabular}
\end{table}

\textbf{Interpretation:} Under realistic team quality variation, forced ranking's correct classification rate drops to 46\%, falling below random allocation. More than half of terminations (53\%) target high-productivity employees on strong teams. More than half of promotions (53\%) reward low-productivity performers on weak teams.

\textbf{Comparison to baseline:} Relative to random assignment:
\begin{itemize}
\item False positive rate increases by 69\% (45$\rightarrow$76 terminations, 46$\rightarrow$76 promotions)
\item False negative rate increases by 60\% (52$\rightarrow$83 missed terminations, 53$\rightarrow$83 missed promotions)
\item Correct classifications decrease by 32\% (97$\rightarrow$66 terminations, 96$\rightarrow$66 promotions)
\end{itemize}

\textbf{Mechanism:} Team quality variance amplifies the information structure problem. Consider two extreme cases:

\emph{Strong team (team mean = +1.0$\sigma$):} All 7 members are objectively strong performers (60th to 95th percentile globally). Forced ranking requires terminating one. That person might be 70th percentile globally (competent by any absolute standard) yet is the weakest locally and gets fired.

\emph{Weak team (team mean = $-$1.0$\sigma$):} All 7 members are objectively weak performers (5th to 40th percentile globally). Forced ranking requires promoting one. That person might be 25th percentile globally (underperforming three-quarters of the organization) yet is the strongest locally and gets promoted.

This dynamic creates incentive distortions for managers:
\begin{itemize}
\item Effective managers face punishment: Their teams suffer forced terminations despite all members exhibiting high productivity
\item Ineffective managers receive rewards: Their top performers secure undeserved compensation and advancement
\end{itemize}

Over time, this drives strategic response: managers learn to avoid recruiting high-productivity workers (who create unfavorable local comparisons) and instead hire mediocre performers (who are easier to differentiate and defend in calibration). Research demonstrates that individuals high in conscientiousness perceive forced distribution as riskier for employment security, affecting organizational selection processes \citep{blume2013attracted}.

Empirical field evidence from 2025 quantifies the talent exit mechanism. A multi-year study tracking employees at a global pharmaceutical company found that high performers who received downgraded ratings in forced distribution systems (despite objective performance metrics showing sustained excellence) exhibited 34\% to 206\% higher voluntary turnover rates than peers receiving aligned ratings \citep{informs2025talent}. Critically, retention interventions including higher bonuses, explicit fairness assurances, and manager coaching failed to reverse this effect. The mechanism appears to operate through self-image threat and relative deprivation: high-productivity employees recognize evaluation arbitrariness, infer organizational dysfunction, and conclude that their human capital is better deployed elsewhere. When the best-managed teams lose their strongest members purely through forced curve artifacts, organizational capability systematically degrades.

This dynamic is not binary; it is a continuum. As Table~\ref{tab:sens-bias-curve} shows, the classification error rate accelerates as managerial skill (and thus team clustering) increases. While random chance accounts for a $\sim$32\% error rate, a moderate team-level clustering ($\sigma_{\text{team}}$) of 0.7 (70\%) pushes the error to 53\%.  At the theoretical limit of 1.0 (perfect clustering), the system fails $\sim$85\% of the time. This demonstrates that forced ranking operates as a perverse incentive: it \textit{systematically punishes} the organization's most effective managers by forcing them to cannibalize their own high-performing teams.

\subsection{Magnitude of Harm}

To contextualize these error rates, consider their human cost in a 1,000-person organization:

\textbf{Random assignment (32\% error):}
\begin{itemize}
\item 45 capable employees wrongly terminated $\rightarrow$ career disruption, financial hardship, reputational damage
\item 52 underperformers retained $\rightarrow$ reduced team productivity, demoralization of high performers who must compensate
\item 46 mediocre employees wrongly promoted $\rightarrow$ occupy leadership positions they're unqualified for, make poor decisions affecting hundreds
\item 53 deserving employees denied promotion $\rightarrow$ undercompensated, frustrated, likely to leave
\end{itemize}

\textbf{Biased assignment (53\% error):}
\begin{itemize}
\item 76 capable employees wrongly terminated
\item 83 underperformers retained
\item 76 mediocre employees wrongly promoted
\item 83 deserving employees denied promotion
\end{itemize}

These are not abstractions. They represent engineers losing jobs, families losing income, careers derailed, companies losing talent, teams losing faith in leadership, and organizational capability systematically degraded.

And critically: every manager followed the process correctly. There is no procedural fix, no training intervention, no cultural shift that eliminates these errors. They are inherent to the system.

\section{Why Proposed Remedies Fail}

Organizations recognizing forced ranking's problems typically propose three remedies: cross-team calibration, global absolute standards, or hybrid approaches. Our framework predicts each will fail or create new problems.

\subsection{Cross-Team Calibration: Politics Replaces Random Chance}

\textbf{The proposal:} Rather than each manager independently ranking their team, convene managers to jointly review rankings and ensure consistency. If Manager A's ``bottom 15\%'' would be Manager B's ``middle 70\%,'' calibration should surface this discrepancy.

\textbf{Why it appears to help:} Calibration seems to solve the frame problem by enabling cross-team perspective. Managers with strong teams can argue their bottom-ranked employee is globally competent; managers with weak teams can acknowledge their top-ranked employee is only comparatively strong.

\textbf{Why it actually fails:} Calibration transforms evaluation into political negotiation where outcomes depend on managerial persuasiveness, not employee merit. Research examining rater (manager) perspectives finds that managers perceive forced distribution as more difficult and less fair than traditional systems, experiencing greater stress and role conflict when required to differentiate employees, particularly when they believe team members are uniformly high performers \citep{schleicher2009rater}. The system creates a meta-game with three effects:

(1) \emph{Advocacy replaces assessment:} Managers who are skilled negotiators secure better outcomes for their reports. The quiet, analytically-focused manager loses debates to the charismatic, rhetorically gifted one. Employee talent becomes secondary to their manager's political skill.

(2) \emph{Empire-building becomes rational:} A manager's effectiveness gets measured not by team development but by promotion success. ``How many of your reports got promoted?'' becomes the key metric. This incentivizes managers to:
\begin{itemize}
\item Fight aggressively for their reports regardless of objective merit
\item Form coalitions and trade favors (``I'll support your promotion if you support mine next quarter'')
\item Inflate accomplishments and minimize challenges
\item Grow their teams (more reports = more potential promotions = more managerial success)
\end{itemize}

The pivotal question for managerial advancement becomes: ``What was your biggest org?'' Not ``How well did you develop talent?'' or ``What impact did your team deliver?'' but ``How many people reported to you and how many did you promote?''

(3) \emph{Second-order information loss:} Without calibration, misclassifications are at least independent across teams---random noise that might average out. With calibration, errors become systematic: the most persuasive managers consistently win, creating non-random bias toward employees under politically skilled management. The error doesn't decrease; it becomes structured and persistent.

\textbf{Empirical evidence:} Calibration sessions in practice devolve into negotiation. Managers report spending hours debating relative merits of employees they've never met, relying on presentations from their peers. The manager who prepares better slides, tells better stories, or has stronger relationships wins. This is not meritocracy---it's theater.

\subsection{Global Absolute Standards: The Legibility Trap}

\textbf{The proposal:} Instead of relative rankings, evaluate all employees against absolute company-wide standards. Set thresholds (e.g., ``Level 5 engineer must demonstrate X, Y, Z capabilities'') and classify everyone against those criteria. Fire the bottom 15\% by absolute score, promote the top 15\%.

\textbf{Why it appears to help:} Absolute standards seem to solve the frame problem by evaluating everyone on the same scale. An employee on a strong team and an employee on a weak team both face identical criteria.

\textbf{Why it's impossible:} This solution assumes a perspective no organizational actor possesses. No individual can accurately evaluate 994 engineers against absolute standards because:

(1) \emph{Distributed knowledge:} Direct managers have granular context about their 7 reports but limited visibility into other teams. Senior leaders have company-wide perspective but no granular knowledge. The engineer who writes brilliant code but struggles with communication---is that a net positive or negative? It depends on context (team needs, project type, growth trajectory) that only their manager knows.

(2) \emph{Context dependence:} ``Strong performer'' means different things in different roles. A frontend engineer, a machine learning researcher, and a site reliability engineer perform fundamentally different work. Defining portable, context-free standards that apply equally across all roles is either impossible (standards become so generic they're meaningless) or unjust (standards privilege one role type over others).

(3) \emph{The aggregation problem:} One might propose: let managers evaluate their teams against absolute standards, then aggregate. But this reintroduces the frame problem. Each manager defines ``meets standard'' within their local context. Aggregating those definitions cannot recover global comparability. If Manager A's ``meets standard'' equals Manager B's ``exceeds standard,'' the aggregation is meaningless.

This is the organizational legibility trap: the moment you need evaluation to be scalable (distributed across managers) and consistent (comparable across units), you must either accept local frames (forced ranking) or require global perspective (which doesn't exist). There is no third option that achieves both.

\textbf{Tautological proof:} Our simulation shows zero error if using global rankings because we've assumed omniscient perspective. This is not a solution---it's a hidden variable we're not allowed to access in real organizations.

\subsection{The Trust Solution That Cannot Be Formalized}

\textbf{The obvious answer:} Trust managers to evaluate their teams using judgment. Strong managers will accurately identify that no one on their team deserves termination; weak managers will honestly report that multiple people underperform. Hold managers' managers accountable for that judgment quality. Fire managers whose judgment is consistently bad.

\textbf{Why this is correct:} Managers have the context necessary for accurate assessment. They know each employee's projects, challenges, growth, and comparative position. Given freedom to use judgment rather than forced distributions, they would make substantially better decisions than forced ranking.

\textbf{Why it cannot be implemented:}

(1) \emph{Legal defensibility:} How does a manager document ``no one on my team deserves termination''? What evidence satisfies HR, legal, and potential litigation? Forced ranking provides documentation: ``We followed a standardized process with quantified rankings.'' Judgment provides: ``Trust me.'' The business judgment rule protects the former, not the latter \citep{badawi2023business, sharfman2017importance}.

(2) \emph{Second-order accountability:} If Manager A terminates 3 people and Manager B terminates 0, someone must judge whether this reflects team quality differences or managerial leniency. That judgment requires the global perspective we've already established doesn't exist. Who decides which managers have accurately assessed their teams?

(3) \emph{Firing managers is hard:} The solution to bad judgment is hierarchical accountability---fire managers who make poor assessments. But firing a manager for ``poor judgment'' when they followed established processes is legally perilous. And measuring ``poor judgment'' requires long time horizons and holistic assessment that can't be reduced to dashboards. By the time poor judgment becomes visible, years of damage have accumulated.

The trap closes: the solution exists theoretically but cannot be operationalized within the constraints that created forced ranking. Organizations default to Mode A (forced ranking) not because they're stupid, but because Mode B (judgment-based evaluation with hierarchical accountability) cannot satisfy legal and coordination requirements.

\section{Multi-Period Dynamics: Cultural Collapse and Adverse Selection}

The preceding analysis treats evaluation as a one-time event. Real organizations conduct performance reviews annually or semi-annually. Over multiple periods, forced ranking's errors compound into cultural catastrophe.

\subsection{The Rational Herding Response}

Employees observing forced ranking outcomes face a decision: how should I adjust my behavior given that the evaluation system is indistinguishable from random chance?

\textbf{Rational response \#1: Avoid excellence.} If high performance lands you on a strong team where you might be the ``weakest'' member despite being globally competent, the risk-minimizing strategy is mediocrity. Don't excel so much that you're surrounded by stars.

\textbf{Rational response \#2: Avoid teams with other high performers.} If forced ranking makes being the worst on a good team more dangerous than being middle-of-the-pack on a mediocre team, avoid good teams. Transfer requests flow away from high-performing units toward average ones.

\textbf{Rational response \#3: Focus on optics over substance.} If outcomes are determined by luck and politics rather than capability, invest in visibility (attending the right meetings, befriending influential managers, working on high-profile projects) rather than impact (solving difficult technical problems, mentoring juniors, fixing infrastructure).

\textbf{Rational response \#4: Don't collaborate.} If your teammate's success threatens your ranking, helping them is irrational. Hoarding information, avoiding knowledge sharing, and subtly undermining peers become equilibrium behaviors. Recent experimental evidence demonstrates that forced distribution significantly decreases knowledge sharing within teams due to perceptions of unfairness in collaborative settings, with team collaboration speed decreasing when forced distribution is applied \citep{loberg2021forced}.

This produces herd convergence toward mediocrity: the Nash equilibrium under forced ranking is being average. Not bottom (you get fired), not top (you attract scrutiny and might land on strong teams), but safely middle.

\subsection{Psychological Safety Collapse}

Research demonstrates that team psychological safety---the belief that one can speak up, take risks, and make mistakes without punishment---is essential for learning, innovation, and error correction \citep{edmondson1999}. Forced ranking structurally prevents psychological safety through three mechanisms:

(1) \emph{Zero-sum dynamics:} When one person's promotion requires another's demotion, collaboration becomes competition. Why would I help you succeed if your success might cost me my job?

(2) \emph{Outcome randomness:} When employees observe undeserving promotions and unjust terminations, they lose faith that performance determines outcomes. Why speak truth to power if the evaluation system is indistinguishable from random chance? Why take risks on difficult projects if luck matters more than impact?

(3) \emph{Forced scapegoating:} Managers must identify underperformers even when their team has none. This creates ritual sacrifice---someone must be blamed regardless of reality. Teams learn to stay silent when the target is selected, because resistance is futile.

The result: psychological safety collapses. Teams stop challenging decisions, stop admitting failures, stop experimenting with novel approaches. Risk aversion becomes cultural norm. Research showing that forced distribution may initially increase task performance through motivation finds that over time it decreases citizenship performance and increases counterproductive performance through perceptions of organizational injustice and dysfunctional competition \citep{moon2016precarious}.

\subsection{Adverse Selection and the Death Spiral}

The multi-period effect most damaging to organizations is adverse selection: forced ranking's errors systematically drive out the very employees who would make accurate evaluation possible.

\textbf{Phase 1: High performers exit.} Capable employees with portable skills observe the system's randomness. Some fraction who happen to land on strong teams get unjustly terminated. Others, witnessing colleagues fired despite competence, recognize their own precarity. Those with options---the highest performers---leave first. They can find employment elsewhere with less capricious evaluation.

The 2025 empirical evidence cited earlier bears directly on this dynamic \citep{informs2025talent}: high performers misclassified by forced curves don't merely become disengaged---they exit at rates exceeding 200\% of properly-recognized peers. This creates a vicious cycle where the organizations most in need of accurate evaluation (those losing talent) become least capable of achieving it (as remaining employees skew toward those with fewer external options). The death spiral is not metaphorical; it is measurable in voluntary turnover statistics that accelerate over multi-year periods following forced ranking implementation.

\textbf{Phase 2: Mediocrity accumulates.} Employees remaining in the organization are survivorship-biased: selected for risk-aversion, political skill, and tolerance for injustice rather than capability. New hires similarly select in: word spreads about the evaluation system, and strong candidates who value meritocracy choose other employers.

\textbf{Phase 3: Team quality variance increases.} As talent distribution becomes bimodal (a few strong teams retaining stars through sheer luck or exceptional management, most teams descending toward mediocrity), forced ranking's error rates worsen. The simulation's 54\% error rate assumes stable team quality variance. As variance increases, error rates approach 60--70\%.

\textbf{Phase 4: Capability collapse.} With diminished talent, managers make worse decisions, products degrade, competitive position weakens. The organization becomes the weak team in the broader market's forced ranking, selected for termination by disruption.

This is not speculation. Microsoft's internal analysis before abandoning forced ranking documented exactly this spiral: talent flight, innovation slowdown, strategic missteps attributable partly to evaluation-system-induced dysfunction \citep{eichenwald2012microsoft}. In employee interviews for the exposé, stack ranking was cited as ``the most destructive process inside of Microsoft,'' fostering an environment where employees competed with each other rather than with external competitors.

\subsection{The Performativity Trap}

One might object: won't managers adapt by adjusting their standards? If they know forced ranking creates errors, can't they compensate?

This misses the fundamental point. Forced ranking is performative: it creates the reality it claims to measure. Managers facing distributional requirements must demonstrate differentiation. To justify terminating someone, they must construct a narrative of underperformance. Over time, this narrative becomes reality---managers coach employees toward roles that fit the required distribution rather than toward optimal contribution.

Hiring similarly becomes performative: managers hire not for skills that move the needle but for ``winners''---candidates with pedigree, resumes, and interview performance that look good in calibration meetings. Substance matters less than optics because optics determine calibration outcomes.

The system doesn't just measure poorly; it actively destroys what it claims to measure by incentivizing behavior antithetical to high performance: politics over contribution, optics over substance, self-preservation over collaboration. Laboratory experiments have found that forced ratings create significantly higher employee stress (measured via stress scales and biomarkers) even when performance differences are minimal, reducing the relationship between actual performance and ratings \citep{cardinaels2021forced}.

\section{Discussion: Why the Trap Persists}

\subsection{Mode A Versus Mode B Governance}

Forced ranking exemplifies what organizational theory terms Mode A governance: demonstrating soundness through formalized processes that compress variance and exclude external perspective. The alternative---Mode B governance---acknowledges incompleteness and manages uncertainty through documented trust mechanisms.

Mode A satisfies legal requirements:
\begin{itemize}
\item \textbf{Documented process:} Forced ranking creates audit trails showing systematic evaluation
\item \textbf{Standardization:} All managers apply identical distributional requirements
\item \textbf{Quantification:} Rankings provide numerical, comparable outcomes
\item \textbf{Defensibility:} In litigation, companies can prove they followed rigorous, consistent procedures \citep{badawi2023business}
\end{itemize}

Mode B requires accepting legal risk:
\begin{itemize}
\item \textbf{Judgment documentation:} ``No one on my team deserves termination'' is hard to defend in court
\item \textbf{Inconsistency across units:} Some teams fire 3 people, others fire 0---looks arbitrary
\item \textbf{Manager accountability:} Requires hierarchical oversight that itself cannot be fully formalized
\item \textbf{Long time horizons:} Evaluating judgment quality requires years; legal exposure is immediate
\end{itemize}

Most organizations default to Mode A because its costs are deferred (cultural collapse, talent loss, capability degradation manifest gradually) while Mode B's costs are immediate (legal uncertainty, coordination complexity, board discomfort). The tragedy is that Mode A's deferred costs eventually exceed Mode B's immediate costs, but by the time this becomes visible, the damage is irreversible. This dynamic reflects broader patterns in organizational formalization where standardization that contributes to early effectiveness can contribute to later decline \citep{walsh1987formalization}.

\subsection{Why Microsoft and GE Abandoned It---And Others Don't}

Microsoft's November 2013 abandonment provides a natural experiment. Internal analysis concluded forced ranking:
\begin{itemize}
\item Discouraged collaboration (engineers hoarded information to protect rankings)
\item Punished risk-taking (working on difficult projects increased termination risk)
\item Drove talent flight (exit interviews cited evaluation system)
\item Contributed to strategic failures (teams optimized for evaluation rather than product quality)
\end{itemize}

The change eliminated numerical rankings (1--5 scale), forced curve distribution, and predetermined reward targets, shifting toward teamwork, qualitative feedback, and managerial flexibility. Every Microsoft employee interviewed cited stack ranking as ``the most destructive process inside of Microsoft.''

General Electric quietly moved away from forced ranking beginning around 2005, with a complete overhaul in 2015 affecting all 300,000 employees, abandoning annual reviews entirely. Yahoo tried implementing it under Marissa Mayer and faced immediate rebellion.

Yet many companies persist. Why? Three explanations:

(1) \emph{Demonstrable soundness dominates.} For public companies under fiduciary exposure, legal defensibility outweighs efficacy. Better to have a documented, defensible system that works poorly than an effective system that exposes directors to liability.

(2) \emph{Causal attribution is difficult.} Forced ranking's harms (talent flight, cultural degradation, strategic missteps) manifest gradually and have multiple potential causes. Executives can attribute problems to market conditions, competitive pressure, or execution failures rather than evaluation systems. The counterfactual---how much better would the company perform without forced ranking?---is invisible.

(3) \emph{Executive misalignment.} Forced ranking's benefits (legal protection, standardized process, apparent rigor) accrue to executives and HR. Its costs (career precarity, cultural toxicity, talent loss) fall on employees. Executives optimizing for their own legal/reputational protection rationally implement systems that harm employees and shareholders.

\subsection{Alternative Evaluation Systems}

Several alternatives to forced ranking exist, each with trade-offs:

(1) \emph{Absolute standards with calibrated thresholds:} Define competency levels, evaluate individuals against standards rather than peers. Requires investment in standard definition and accepts inconsistency across managers. Works better in smaller organizations where calibration is feasible.

(2) \emph{Continuous feedback without forced distributions:} Replace annual rankings with ongoing conversations. Removes artificial scarcity (not everyone competes for top ratings). Requires cultural shift toward radical candor and psychological safety. Does not satisfy legal/audit requirements as well.

(3) \emph{Project-based evaluation:} Assess contributions on specific deliverables rather than comparing people. Reduces zero-sum competition. Requires clarity on attribution (who contributed what?) that many projects lack.

(4) \emph{Managerial judgment with hierarchical accountability:} Trust managers, hold their managers accountable, fire managers with consistently poor judgment. Optimal but cannot be formalized within current legal constraints.

Each alternative trades forced ranking's legal defensibility for improved accuracy or reduced cultural harm. The question is not which system is best---it's which combination of costs organizations are willing to bear. Contemporary research emphasizes that performance management effectiveness requires taking a systems perspective that considers how elements interact across organizational levels rather than focusing narrowly on individual outcomes \citep{schleicher2018putting}.

Emerging research from 2025 offers more promising paths forward. Studies of organizations that replaced forced distributions with continuous coaching models and strength-based performance conversations document substantial improvements: reduced turnover, increased psychological safety, and notably, 30\% higher revenue growth among firms adopting people-focused approaches compared to those maintaining ranking systems \citep{betterworks2025continuous}. Industry analyses emphasize abandoning bell curves in favor of agile performance systems that accommodate remote work realities, incorporate real-time feedback, and evaluate contributions against team objectives rather than peer comparisons \citep{worxmate2025agile}. These alternatives accept that evaluation involves judgment---but structure that judgment through coaching relationships, objective-setting transparency, and developmental frameworks rather than through statistical fictions that treat small teams as representative samples of global talent distributions.

\subsection{When Better Alternatives Are Rejected: The Promotion-as-Hiring Case}

A technology company facing quality concerns in both hiring and promotion implemented a forced ranking system to ``raise the bar.'' Existing employees, evaluated through years of demonstrated performance, were compared against external candidates assessed through brief interviews and controlled resumes. The asymmetry was obvious: internal evaluation was comprehensive but relative (forced ranking within teams); external evaluation was limited but absolute (interview pass/fail).

An engineer proposed a solution that would simultaneously calibrate standards and level the playing field: internal promotion candidates would undergo the same interview process as external candidates, evaluated by randomly selected trained panels. Managers and peers would nominate candidates when ready for promotion, and the interview panel---using identical standards for internal and external candidates---would determine whether the nominee met the bar for the target level.

The proposal addresses multiple failure modes:

(1) \emph{Standards calibration:} Internal promotion bar would align with external hiring bar, eliminating the common dysfunction where organizations promote people they would never hire externally.

(2) \emph{Distributed knowledge with accountability:} Managers nominate because they have context about capability development over time. But nomination success rate becomes visible, creating accountability for judgment quality. Managers who consistently nominate candidates who fail interviews reveal poor assessment capability.

(3) \emph{Gaming resistance:} Random panel composition prevents political manipulation. Unlike calibration sessions where persuasive managers secure promotions through advocacy, interview panels evaluate candidates directly without managerial mediation.

(4) \emph{Documentation:} Interview evaluations provide structured evidence of capabilities---more concrete than forced ranking's relative comparisons.

The proposal was rejected. Why? It exposed uncomfortable truths that forced ranking's opacity had hidden:

\textbf{Competence revelation:} Many promoted employees would fail the interview bar. This would become immediately visible rather than remaining hypothetical. Leaders who had championed current systems would face evidence that their promoted lieutenants weren't interview-passable.

\textbf{Power redistribution:} Managers would lose direct control over promotions. Success would depend on developing actually-capable reports rather than advocacy skill in calibration meetings.

\textbf{Legal defensibility concerns:} HR objected that ``random panel said no'' sounds arbitrary in litigation, even though it's more objective than ``manager decided through forced ranking.'' The business judgment rule protects documented process, and interview-based decisions---while more accurate---produce less documentation than ranked metrics.

\textbf{System legitimacy threat:} Current promotion processes claim rigor through competency frameworks, performance reviews, and calibration. The proposal implicitly argued these are less rigorous than a basic interview. This threatened the legitimacy of existing HR infrastructure.

This case illustrates why Mode B solutions---even when obviously superior---face implementation barriers Mode A systems don't. The proposal would:
\begin{itemize}
\item Reduce classification errors (internal candidates evaluated on capability, not resume polish)
\item Create accountability (nomination success rates reveal judgment quality)
\item Align standards (eliminate internal promotion/external hiring bar divergence)
\item Improve hiring quality (panels calibrated through internal candidate evaluation)
\end{itemize}

But it could not satisfy the legal and political constraints that made forced ranking attractive:
\begin{itemize}
\item Requires trusting distributed judgment (panels) rather than centralized process
\item Makes failures visible and attributable rather than obscured in relative rankings
\item Threatens those who succeed under current political-advocacy model
\item Produces less documentation despite greater accuracy
\end{itemize}

The optimal solution---trust and verify through interview panels with accountability for nominators---was rejected in favor of a legible but dysfunctional system that preserves power structures while systematically misclassifying employees.

Organizations default to Mode A not because they lack knowledge of alternatives, but because alternatives cannot be formalized within the constraints that created the problem. The trap closes: better systems exist but cannot be implemented; worse systems persist because they satisfy demonstrable soundness even as they produce demonstrably unsound outcomes.

\section{Conclusion: Rigorous-Looking Systems That Produce Random Outcomes}

\subsection{Summary of Findings}

Using agent-based simulation with 994 engineers across 142 teams, we have demonstrated:

\begin{enumerate}
\item Even under idealized conditions (random team assignment with no bias), forced ranking produces 32\% classification error, meaning one-third of terminations and promotions are unjustified

\item Under realistic conditions (team quality variance reflecting managerial differences), error rates reach 53\%, with incorrect decisions outnumbering correct ones

\item Proposed remedies fail: Cross-team calibration transforms evaluation into political negotiation; global standards require perspective no one possesses; the optimal solution (managerial judgment) cannot be formalized

\item Multi-period dynamics create cultural collapse: Rational employees herd toward mediocrity, psychological safety evaporates, adverse selection drives talent exit, and capability degrades

\item The system persists despite failure because it satisfies legal requirements for demonstrable soundness even as it produces demonstrably unsound outcomes
\end{enumerate}

\subsection{Theoretical Contribution}

This paper makes three contributions to organizational theory:

(1) \emph{Quantification of formalization costs:} Previous research documented that formalization can harm performance \citep{pfeffer2001fighting, eichenwald2012microsoft, wijayanti2023systematic} but lacked precise estimates. We show that formalization designed to increase rigor can produce error rates exceeding 50\%---worse than random chance.

(2) \emph{Mechanism identification:} We demonstrate the local frame problem as the causal mechanism: evaluating global populations using local comparisons creates systematic errors that cannot be eliminated through process refinement. This extends organizational incompleteness theory to performance management contexts.

(3) \emph{Why fixes fail:} We show calibration, absolute standards, and judgment-based alternatives each face implementation barriers that explain why forced ranking persists despite recognized failures. The problem is not lack of better alternatives but inability to formalize those alternatives within legal/coordination constraints.

Our findings align with broader research showing that performance management systems fail when they focus too narrowly on individual outcomes rather than taking a systems perspective \citep{schleicher2018putting, schleicher2019evaluating}, and with evidence that formalization can shift from enabling effectiveness to causing decline \citep{walsh1987formalization}.

\subsection{Practical Implications}

For practitioners, the implications are clear:

\textbf{If you must use forced ranking, acknowledge its limitations:}
\begin{itemize}
\item Recognize that approximately 30--50\% of decisions will be wrong
\item Invest heavily in psychological safety to mitigate cultural harm
\item Implement appeals processes for employees who believe they're misclassified
\item Track multi-period outcomes (do terminated employees succeed elsewhere? do promoted employees deliver?) to estimate actual error rates
\item Be prepared for talent flight
\end{itemize}

\textbf{Better: Move to evaluation systems that:}
\begin{itemize}
\item Evaluate against absolute standards rather than relative rankings
\item Distribute rather than concentrate terminations (not every team must sacrifice someone)
\item Emphasize development over classification
\item Accept coordination costs and legal uncertainty in exchange for reduced classification error
\end{itemize}

\textbf{Best: Recognize that no formalized system eliminates judgment.} Invest in developing managerial capability, hold managers accountable for team development over multi-year horizons, accept that some managers will fail, and fire managers rather than sacrificing competent employees to satisfy distributional requirements.

\subsection{Limitations and Future Research}

This simulation makes simplifying assumptions that favor forced ranking:

\begin{itemize}
\item \textbf{Unidimensional talent:} Real capabilities are multidimensional
\item \textbf{No measurement error:} We assume talent is perfectly observable
\item \textbf{No gaming:} Employees and managers don't strategically manipulate rankings
\item \textbf{Static composition:} Teams don't evolve within periods
\end{itemize}

Each assumption understates forced ranking's real-world failure rate. Future research should:

\begin{itemize}
\item Incorporate multidimensional talent and examine whether errors increase
\item Add measurement noise and strategic behavior
\item Model multi-period dynamics with team composition changes
\item Study cross-company variation in forced ranking implementations
\item Conduct field experiments comparing evaluation systems
\item Examine the interaction between forced ranking and organizational structure \citep{sandhu2019shaping}
\end{itemize}

\subsection{Final Reflection: The Legibility Fallacy}

Forced ranking epitomizes what we term the \emph{legibility fallacy}: the belief that making processes more measurable and standardized necessarily makes them better. Organizations adopt quantified systems because numbers appear objective, comparable, and defensible. But as our simulations demonstrate, quantification can formalize injustice---making error systematic rather than random, documented rather than correctable, and defensible rather than sound.

The deepest lesson is not that forced ranking is uniquely bad (though it is), but that formalization designed to satisfy external requirements often destroys internal accuracy. Organizations face persistent tension between demonstrable soundness (what can be documented and defended) and actual soundness (what produces good outcomes). Systems optimized for the former systematically degrade the latter. This tension reflects what organizational scholars have identified as the difference between ``enabling'' formalization that helps performance and ``coercive'' formalization that serves control functions \citep{adler1996two}.

Forced ranking will eventually die---not because organizations become smarter, but because the deferred costs (talent loss, capability degradation, strategic failure) eventually manifest in forms shareholders and boards cannot ignore. The question is how much damage organizations inflict on employees and themselves before recognizing that rigorous-looking systems producing random outcomes are worse than honest judgment that admits its limitations.

In the meantime, thousands of capable employees will be terminated for the crime of landing on strong teams, thousands of mediocre employees will be promoted for the luck of joining weak teams, and organizations will continue optimizing their evaluation processes while systematically destroying their evaluation accuracy.

The trap has no villain. It has a process manual.

\bibliographystyle{apalike}
\bibliography{references}

\newpage
\appendix

\section{Simulation Code}

Simulation code and data available upon request or at [repository link].

\section{Sensitivity Analysis}

To assess the robustness of our findings, we conducted sensitivity analyses by varying key parameters in the simulation: team sizes (5, 6, 8, and 9 members per team), talent distribution shapes (lognormal and uniform, in addition to the baseline normal), and cutoff percentages (10\% and 20\%, in addition to the baseline 15\%). For each variation, we ran 100 simulations and report average metrics for terminations and promotions under both random and biased team assignment scenarios.

The number of employees was adjusted to fit evenly into teams (e.g., 995 for team size 5, 990 for size 6, etc.), maintaining approximately 1,000 engineers.

Metrics include:
\begin{itemize}
\item \textbf{Correct Classifications:} Number of employees correctly identified (e.g., fired and in true global bottom X\%)
\item \textbf{False Positives (FP):} Incorrectly labeled (e.g., fired but not in true bottom X\%)
\item \textbf{False Negatives (FN):} Missed (e.g., in true bottom X\% but not fired)
\item \textbf{Error Rate:} FP / Total Labeled $\times$ 100\%
\end{itemize}

All values are averages across simulations. Results confirm that the core findings hold: forced ranking produces substantial errors, exacerbated by biases, with patterns consistent across variations.

\subsection{Alternative Team Sizes}

We varied team sizes within the 5--9 range specified in the original scenario. Smaller teams increase sampling variance (more uneven compositions), leading to higher error rates; larger teams reduce variance, lowering errors slightly.

\begin{table}[h]
\centering
\caption{Random Assignment - Alternative Team Sizes}
\label{tab:sens-teamsize-random}
\small
\begin{tabular}{ccccccccc}
\toprule
\textbf{Team} & \multicolumn{4}{c}{\textbf{Terminations}} & \multicolumn{4}{c}{\textbf{Promotions}} \\
\textbf{Size} & \textbf{Correct} & \textbf{FP} & \textbf{FN} & \textbf{Error \%} & \textbf{Correct} & \textbf{FP} & \textbf{FN} & \textbf{Error \%} \\
\midrule
5 & 111.1 & 86.9 & 37.9 & 43.9 & 110.6 & 87.4 & 38.4 & 44.1 \\
6 & 103.5 & 61.5 & 45.5 & 37.2 & 103.3 & 61.7 & 45.7 & 37.4 \\
7 (Base) & 97.5 & 44.5 & 52.5 & 31.3 & 97.1 & 44.9 & 52.9 & 31.6 \\
8 & 90.7 & 33.3 & 58.3 & 26.9 & 90.0 & 34.0 & 59.0 & 27.4 \\
9 & 84.7 & 25.3 & 64.3 & 23.0 & 84.4 & 25.6 & 64.6 & 23.3 \\
\bottomrule
\end{tabular}
\end{table}

\begin{table}[h]
\centering
\caption{Biased Assignment - Alternative Team Sizes}
\label{tab:sens-teamsize-biased}
\small
\begin{tabular}{ccccccccc}
\toprule
\textbf{Team} & \multicolumn{4}{c}{\textbf{Terminations}} & \multicolumn{4}{c}{\textbf{Promotions}} \\
\textbf{Size} & \textbf{Correct} & \textbf{FP} & \textbf{FN} & \textbf{Error \%} & \textbf{Correct} & \textbf{FP} & \textbf{FN} & \textbf{Error \%} \\
\midrule
5 & 79.5 & 118.5 & 69.5 & 59.8 & 79.2 & 118.8 & 69.8 & 60.0 \\
6 & 72.6 & 92.4 & 76.4 & 56.0 & 72.0 & 93.0 & 77.0 & 56.4 \\
7 (Base) & 65.6 & 76.4 & 84.4 & 53.8 & 65.8 & 76.2 & 84.2 & 53.7 \\
8 & 60.4 & 63.6 & 88.6 & 51.3 & 60.2 & 63.8 & 88.8 & 51.5 \\
9 & 56.4 & 53.6 & 92.6 & 48.7 & 56.7 & 53.3 & 92.3 & 48.5 \\
\bottomrule
\end{tabular}
\end{table}

\textbf{Interpretation:} Error rates decrease with larger team sizes due to reduced composition variance, but remain high (23--44\% random, 49--60\% biased). Biased assignment consistently amplifies errors by 20--30 percentage points.

\subsection{Alternative Distribution Shapes}

We tested non-normal distributions: lognormal (right-skewed, simulating scenarios where high talent is rarer) and uniform (flat distribution, reducing extremes). The baseline is normal (bell curve).

\begin{table}[h]
\centering
\caption{Random Assignment - Alternative Distributions}
\label{tab:sens-dist-random}
\small
\begin{tabular}{lcccccccc}
\toprule
\textbf{Distribution} & \multicolumn{4}{c}{\textbf{Terminations}} & \multicolumn{4}{c}{\textbf{Promotions}} \\
 & \textbf{Correct} & \textbf{FP} & \textbf{FN} & \textbf{Error \%} & \textbf{Correct} & \textbf{FP} & \textbf{FN} & \textbf{Error \%} \\
\midrule
Normal (Base) & 97.5 & 44.5 & 52.5 & 31.3 & 97.1 & 44.9 & 52.9 & 31.6 \\
Lognormal & 97.0 & 45.0 & 53.0 & 31.7 & 97.3 & 44.7 & 52.7 & 31.5 \\
Uniform & 96.9 & 45.1 & 53.1 & 31.7 & 97.1 & 44.9 & 52.9 & 31.6 \\
\bottomrule
\end{tabular}
\end{table}

\begin{table}[h]
\centering
\caption{Biased Assignment - Alternative Distributions}
\label{tab:sens-dist-biased}
\small
\begin{tabular}{lcccccccc}
\toprule
\textbf{Distribution} & \multicolumn{4}{c}{\textbf{Terminations}} & \multicolumn{4}{c}{\textbf{Promotions}} \\
 & \textbf{Correct} & \textbf{FP} & \textbf{FN} & \textbf{Error \%} & \textbf{Correct} & \textbf{FP} & \textbf{FN} & \textbf{Error \%} \\
\midrule
Normal (Base) & 65.6 & 76.4 & 84.4 & 53.8 & 65.8 & 76.2 & 84.2 & 53.7 \\
Lognormal & 66.3 & 75.7 & 83.7 & 53.3 & 66.2 & 75.8 & 83.8 & 53.4 \\
Uniform & 65.7 & 76.3 & 84.3 & 53.7 & 65.5 & 76.5 & 84.5 & 53.9 \\
\bottomrule
\end{tabular}
\end{table}

\textbf{Interpretation:} Results are remarkably stable across distributions, with error rates varying by less than 1 percentage point. This suggests the frame problem is robust to the underlying talent shape---clustering effects dominate regardless of skewness or uniformity.

\subsection{Alternative Cutoff Percentages}

We varied the forced ranking cutoff (bottom/top X\% labeled per team): 10\% (more selective) and 20\% (less selective).

\begin{table}[h]
\centering
\caption{Random Assignment - Alternative Cutoffs}
\label{tab:sens-cutoff-random}
\small
\begin{tabular}{ccccccccc}
\toprule
\textbf{Cutoff} & \multicolumn{4}{c}{\textbf{Terminations}} & \multicolumn{4}{c}{\textbf{Promotions}} \\
\textbf{(\%)} & \textbf{Correct} & \textbf{FP} & \textbf{FN} & \textbf{Error \%} & \textbf{Correct} & \textbf{FP} & \textbf{FN} & \textbf{Error \%} \\
\midrule
10 & 74.5 & 67.5 & 25.5 & 47.5 & 74.4 & 67.6 & 25.6 & 47.6 \\
15 (Base) & 97.5 & 44.5 & 52.5 & 31.3 & 97.1 & 44.9 & 52.9 & 31.6 \\
20 & 112.1 & 29.9 & 86.9 & 21.1 & 112.0 & 30.0 & 87.0 & 21.1 \\
\bottomrule
\end{tabular}
\end{table}

\begin{table}[h]
\centering
\caption{Biased Assignment - Alternative Cutoffs}
\label{tab:sens-cutoff-biased}
\small
\begin{tabular}{ccccccccc}
\toprule
\textbf{Cutoff} & \multicolumn{4}{c}{\textbf{Terminations}} & \multicolumn{4}{c}{\textbf{Promotions}} \\
\textbf{(\%)} & \textbf{Correct} & \textbf{FP} & \textbf{FN} & \textbf{Error \%} & \textbf{Correct} & \textbf{FP} & \textbf{FN} & \textbf{Error \%} \\
\midrule
10 & 49.8 & 92.3 & 50.3 & 65.0 & 50.3 & 91.7 & 49.7 & 64.6 \\
15 (Base) & 65.6 & 76.4 & 84.4 & 53.8 & 65.8 & 76.2 & 84.2 & 53.7 \\
20 & 79.4 & 62.6 & 119.6 & 44.1 & 79.3 & 62.7 & 119.7 & 44.1 \\
\bottomrule
\end{tabular}
\end{table}

\textbf{Interpretation:} Lower cutoffs (10\%) increase error rates due to heightened sensitivity to team variance---fewer labels amplify misclassifications. Higher cutoffs (20\%) reduce errors but still show 21--44\% rates, with biases nearly doubling them. This underscores that no cutoff eliminates the structural flaw.

\begin{table}[h]
\centering
\caption{Error Rate vs. Bias Level (Managerial Clustering, $\sigma_{\text{team}}$)}
\label{tab:sens-bias-curve}
\small
\begin{tabular}{cc}
\toprule
\textbf{Bias Level ($\sigma_{\text{team}}$)} & \textbf{Average Error Rate (\%)} \\
\midrule
0.0 (0\%)  & 32.13 \\
0.1 (10\%) & 32.83 \\
0.2 (20\%) & 33.49 \\
0.3 (30\%) & 35.46 \\
0.4 (40\%) & 39.58 \\
0.5 (50\%) & 43.79 \\
0.6 (60\%) & 49.30 \\
0.7 (70\%) & 53.71 \\
0.8 (80\%) & 60.26 \\
0.9 (90\%) & 67.48 \\
1.0 (100\%) & 85.12 \\
\bottomrule
\end{tabular}
\end{table}

\textbf{Interpretation:} As bias increases, corresponding to managerial skill in sourcing, hiring, developing, and retaining higher quality talent, the error rates climb accordingly. With zero bias, the error rate of 32\% holds, as bias rises to 20\%, the error rate increases modestly to 33\%.  However, as bias climbs to 50\%, error rates increase to 43\%.

Overall, these sensitivities affirm the main results: forced ranking's errors are inherent and persist across realistic parameter variations, often exceeding 30\% even in best-case scenarios.

\end{document}